# Aging Studies for the Large Honeycomb Drift Tube System of the Outer Tracker of HERA-B

The HERA-B Outer Tracker Group

H.Albrecht[h], Th.S.Bauer[a,r], M.Beck[q], A.Belkov[g], K.Berkhan[s], G.Bohm[b], M.Bruinsma[a,r], T.Buran[p], M.Capeans[h], J.Chamanina[o], B.X.Chen[d], H.Deckers[e], K.Dehmelt[h], X.Dong[c], R.Eckmann[b], D.Emelianov[h], S.Fourletov[o], I.Golutvin[g], M.Hohlmann[h], K.Höpfner[h], W.Hulsbergen[a], Y.Jia[c], C.Jiang[c], H.Kapitza[f], S.Karabekyan[h,v], Z.Ke[c], Y.Kiryushin[g], H.Kolanoski[e], S.Korpar[k,m], P.Križan[k,l], D.Krücker[e], A.Lanyov[g], Y.Q.Liu[d], T.Lohse[e], R.Loke[e], R.Mankel[e], G.Medin[e], E.Michel[h], A.Moshkin[g], J.Ni[d], S.Nowak[s], M.Ouchrif[a,r], C.Padilla[h], D.Pose[g,j], D.Ressing[h], V.Saveliev[o], B.Schmidt[h], W.Schmidt-Parzefall[i], A.Schreiner[s], U.Schwanke[s], A.S.Schwarz[h], I.Siccama[h], S.Solunin[g], S.Somov[h,t], V.Souvorov[s], A.Spiridonov[s,n], M.Starič[k], C.Stegmann[s], O.Steinkamp[a], N.Tesch[h], I.Tsakov[u], U.Uwer[e,j], S.Vassiliev[g], I.Vukotic[e], M.Walter[s], J.J.Wang[d], Y.M.Wang[d], R.Wurth[h], J.Yang[d], Z.Zheng[c], Z.Zhu[c], R.Zimmermann[q]

[a] *NIKHEF, 1009 DB Amsterdam, The Netherlands* [1]

[b] *University of Texas at Austin, Department of Physics, Austin, TX 78712-1081, USA* [2]

[c] *Institute of High Energy Physics, Beijing 100039, P.R. China*

[d] *Institute of Engineering Physics, Tsinghua University, Beijing 100084, P.R. China*

[e] *Institut für Physik, Humboldt-Universität zu Berlin, D-10115 Berlin, Germany* [3]

[f] *Institut für Physik, Universität Dortmund, D-44227, Germany* [3]

[g] *Joint Institute for Nuclear Research, Dubna, RU-141980, Russia*

[h] *DESY Hamburg, D-22607 Hamburg, Germany*

[i] *Institut für Experimentalphysik, Universität Hamburg, D-22761 Hamburg, Germany* [3]

[j] *Physikalisches Institut, Universität Heidelberg, D-69120 Heidelberg, Germany* [3]

[k] *Jozef Stefan Institute, Experimental Particle Physics Department, 1001 Ljubljana, Slovenia*

[l] *University of Ljubljana, 1001 Ljubljana, Slovenia*

[m] *University of Maribor, 2000 Maribor, Slovenia*

[n] *Institute of Theoretical and Experimental Physics, 117259 Moscow, Russia* [4]

[o] *Moscow Physical Engineering Institute, 115409 Moscow, Russia*

[p] *Department of Physics, University of Oslo, N-0316 Oslo, Norway* [5]

[q] *Fachbereich Physik, Universität Rostock, D-18055 Rostock, Germany* [3]

[r] *Universiteit Utrecht/NIKHEF, 3584 CB Utrecht, The Netherlands* [1]

[s] *DESY Zeuthen, D-15738 Zeuthen, Germany*

[t] *visitor from Moscow Physical Engineering Institute, 115409 Moscow, Russia*

[u] *visitor from Institute for Nuclear Research, INRNE-BAS, Sofia, Bulgaria*

[v] *visitor from Yerevan Physics Institute, Yerevan, Armenia*

[1] *supported by the Foundation for Fundamental Research on Matter (FOM), 3502 GA Utrecht, The Netherlands*

[2] *supported by the U.S. Department of Energy (DOE)*

[3] *supported by the Bundesministerium für Bildung und Forschung, FRG, under contract numbers 05-7BU35I, 05 HB1KHA, 05-7D055P, 05 HB1HRA, 05 HB9HRA, 05 7HD15I, 05 7HH25I*

[4] *supported by the Russion Fundamental Research Foundation under grant RFFI-01-02-17298A and the BMBF via the Max Planck Research Award*

[5] *supported by the Norwegian Research Council*

**Abstract**

The HERA-B Outer Tracker consists of drift tubes folded from polycarbonate foil and is operated with $Ar/CF_4/CO_2$ as drift gas. The detector has to stand radiation levels which are similar to LHC conditions. The first prototypes exposed to radiation in HERA-B suffered severe radiation damage due to the development of self-sustaining currents (Malter effect). In a subsequent extended R&D program major changes to the original concept for the drift tubes (surface conductivity, drift gas, production materials) have been developed and validated for use in harsh radiation environments. In the test program various aging effects (like Malter currents, gain loss due to anode aging and etching of the anode gold surface) have been observed and cures by tuning of operation parameters have been developed.

## 1. Introduction

HERA-B [1] is an experiment designed to measure CP violation in the B system using the so-called gold-plated decay channel, $B^0 \rightarrow J/\psi K^0_S \rightarrow l^+l^-\pi^+\pi^-$ ($l$=e,μ). Events are produced by colliding 920 GeV energy protons accelerated in the HERA proton ring with movable target wires installed in the proton beam halo. In order to accumulate enough statistics within reasonable time, the experiment needs a very high proton-nucleus (pN) interaction rate, and very sophisticated trigger and DAQ systems [2], able to select and fully reconstruct the $B^0 \rightarrow J/\psi K^0_S$ events in a background twelve orders of magnitude larger than the signal.

The First Level Trigger [3] is designed to select events containing two lepton tracks forming an invariant mass compatible to the one of the J/ψ. The two leptons must be identified among an average of 200 charged tracks produced per bunch crossing in the detector when running at 40 MHz interaction rate. The particle density follows an $1/R^2$ dependence, with R being the radial distance to the proton beam. All this requires tracking detectors with very high efficiency, good hit resolution and covering a large area, which is accomplished by using different granularities and technologies: a Silicon Vertex detector near the interaction point [4], Micro Strip Gas Chambers (Inner Tracker) [5] in the forward region closest to the beam pipe, and single-wire drift tubes (Outer Tracker, OTR) to cover the largest area starting at a radius of 19 cm from the beam and extending to an acceptance of 250 mrad. Leptons are identified in the sampling Electromagnetic Calorimeter with shashlik geometry [6] or in the Muon System [7], that uses drift tubes and pad chambers for the outermost region and pixel chambers in the region closest to the beam. Additionally, a RICH detector [8] is used for the π/K separation.

The high event rate needed for the challenging CP violation measurement results in a high radiation level that the experiment has to withstand. This led to several aging problems with the gaseous detectors, which are described in [9,10,11] for components other than the Outer Tracker. In the following, the design constraints and technological choices made after extensive R&D of aging phenomena for the Outer Tracker (OTR) are described in detail. Earlier reports can be found in [12,13,14]. In spite of the conventional technology, the OTR can be considered as a new generation detector that will suffer experimental conditions very similar to those expected at the LHC detectors.

## 2. The HERA-B Outer Tracker

The acceptance coverage, the First Level Trigger requirements, and particle recognition needs of HERA-B are met by a tracker system composed of thirteen measurement Super Layers (SL) of different sizes. SL are split into two halves. Every half SL consists of plane detector layers that are embedded in a common gas box. The largest station has an area of 6.5 x 4.6 $m^2$. Every detector layer consists of detector modules made with honeycomb drift cell technology. The OTR consists of a total of 978 modules and 115000 readout channels. The total cathode surface is 8000 $m^2$ and the total gas volume is 22 $m^3$. In the innermost part, closer to the beam pipe, modules are made with cells of 5 mm diameter. In the outer part, where the track density is lower, 10 mm cells provide the needed coverage minimizing the channel count. Seven stations are installed inside a magnetic field of up to 0.8 Tesla, providing a field integral of 2.1 Tm. The design goals of the system were to provide a hit resolution of about 200 μm, a hit efficiency of 98% and a momentum resolution of $\Delta p/p^2 \sim 10^{-4}$/GeV/c.

The construction scheme of every detector module is shown in fig. 1. A pre-folded, soot-loaded polycarbonate foil of 75 μm thickness (trademark Pokalon-C) is fixed in an aluminium template (not shown in the figure) that defines the hexagonal cell geometry. The foil has the needed bulk conductivity to transport the ion current. One meter long foils are placed along the template to produce modules of up to 4.6 m length. Foils in subsequent layers are staggered by 50 cm. Supporting strips made of FR4 material, 100 μm thick, with metal soldering pads are glued to the foil. These supporting strips are placed every 50-60 cm to reduce the free wire length. At both ends of the module, hexagonal pieces made of NORYL are also glued to the foil to connect the thin cell body with the fixation plate. Gold-plated tungsten wires (6% gold in weight) of 25 μm diameter are strung, tensioned at 50 grams, and soldered to the metal pads on the supporting strips while the cell is still open. To limit the wire occupancy, the cells of modules close to the beam are longitudinally segmented, the shortest segment being 20 cm long. The signals from the inner sectors are transported by 75 μm Cu/Be wires (strung with the same tension and soldered to the support strips) to the module ends. These thicker wires do not give rise to electrostatic fields large enough for significant gas gain. Soldering points are not cleaned up of colophony. Another set of foils is prepared on another template, where glue is applied to the foils. Along the cell, also a dot of conductive glue is applied every 20 cm. The template is then installed on top of the wired half-cells thus forming the hexagonal tube cells. The templates are covered with anticontact paste QZ5111 (from Ciba Spezialitäten GmbH) to avoid that foils stick to them. Complete modules are constructed by repeating the procedure [15]. Since the foil material is not strong enough mechanically to form a self-supporting structure, stabilization carbon rods of 2 mm diameter need to be glued to the external module foils in order to guarantee straightness when modules are assembled in the final vertical position. Electronic boards to provide positive high voltage to the module wires and manufactured with SMD components are installed at the module ends. These boards are inside the gas volume after detector installation.

Signals are read out by front-end amplifier-shaper-discriminator boards (ASD-8) [16] placed at the module end outside the gas box away from the beam pipe where the radiation is below 50 Gy/year. LVDS signals from the preamplifier outputs are digitized in a TDC [17] board, whose information is then sent to the HERA-B readout system.

## 3. Implications of the design for aging studies

The size, running environments and building technology pose some constraints that can affect the aging behaviour of the detector:

- The choice of honeycomb geometry and the construction technique forces the use of a foldable foil (Pokalon-C) and supporting strips (FR4) to fix the wires. Soldering points and glues are inside the amplification region.
- The maximum drift distance in the detector is 5 mm and some of the chambers are inside a magnetic field of up to 0.8 Tesla. In addition, some of the signals have to be transmitted up to 2.2 m. However, in order to trigger signals within one bunch crossing to limit occupancy, the maximum allowed drift time is 75 ns, which is given by the time between proton bunches in HERA plus signal propagation times, TDC delays, etc.. This requires a gas with very high electron drift velocities in the order of 100 μm/ns, which forces operation with a $CF_4$-based gas. Mixtures of $CF_4/CH_4$ (80:20), $Ar/CF_4/CH_4$ (74:20:6) and $Ar/CF_4/CO_2$ (65:30:5) have been tested.
- By design, the drift tubes are not separately gas tight. Gas is distributed to the cells by installing the modules inside a large gas box that is flushed with a gas inlet at the bottom and outlet at the top. Thus, gas is in contact, not only with all the materials of the module detector itself, but also with the gas box and the electronics boards that serve the high voltage. The large gas volume of 22 $m^3$ and the expensive $CF_4$-based gas require the use of a recirculating gas system [18]. Purifiers and outgassing effects of all materials inside the gas box have to be carefully tested.
- The HERA proton beam has an inter-bunch time of 96 ns. To record the needed number of gold-plated $J/\psi K^0_S$ events, the HERA-B target is positioned into the proton beam to produce an average of four interactions per bunch crossing. As shown in fig. 2, under these circumstances, the primary charged particle flux received at HERA-B is equivalent to those that will be seen in detectors at LHC. The observed particle flux, which includes also particles from secondary interactions and gamma conversions, was found to reach (at a distance of 20 cm to the beam) $2 \cdot 10^5$ $cm^{-2}s^{-1}$.

- Due to electronics noise in the front-end electronics, the threshold cannot be lower than about 2.5 fC. On the other hand, the attachment in the $CF_4$-based gas mixtures used is in the order of 3-4, so that the detector has to operate at a gain of about $3 \cdot 10^4$. Consequently, the expected accumulated charge in the hottest areas of the detectors was estimated to reach 0.4-0.5 C/cm per year[1].

Since HERA-B was designed for several years operation, detector and materials had to be tested up to a minimum irradiation of 2-3 C/cm. Outgassing and aging effects of all the materials in the gas box and gas system purifiers had to be carefully tested for such conditions.

## 4. Observation of Malter effect in early prototypes

In first aging studies, small test chambers of honeycomb drift tubes operated with a drift gas mixture of $CF_4/CH_4$ (80:20) were irradiated with X-rays from a tube with a Molybdenum anode (35 kV) up to an accumulated charge of 4.5 C/cm. No aging effects were observed in these tests.

Table 1. Materials used in the first prototype modules tested in HERA-B

| Component | Material |
|---|---|
| Cathode | Soot-loaded Pokalon foil (Pokalon-C) |
| Anode wire | 25 µm Au/W from California Fine Wires |
| Signal transport wire | 75 µm Cu/Be from Little Falls Alloy |
| Supporting wtrips | FR4 with Cu metal pads |
| Supporting glue | Araldite AW 106 and hardener HV953U from Ciba-Geigy AG |
| Conductive glue | Silber Leitkleber 3025 from Epoxy Produkte GmbH |
| Solder tin | Fluitin SN 60% Pb 38% Cu 2% DIN 1707 F-SW26 DIN8516 2.2% flux from Küppers Metallwerk GmbH |
| End-pieces | NORYL |
| Gas box | Aluminium frames and windows |
| Gas system | Cu piping, open loop, water bubbler on exit. |

In 1997, 1-meter long module prototypes produced with the same techniques as described above and the materials listed in Table 1 and operated with $CF_4/CH_4$ (80:20) gas mixture were installed in HERA-B. The chambers were operated with target rates up to 20 MHz. After 3-4 hours of operation, with an accumulated charge of 5-10 mC/cm per wire, the measured chamber currents started to raise reaching values five times larger than at irradiation startup. After removing the target, ie. stopping the irradiation, a rest current remained and only disappeared after switching off the HV. However, once observed, the behavior was triggered again almost immediately under radiation. The same effect was reproduced operating the detector with an Ar/$CF_4$/$CH_4$ (74:20:6) gas mixture. Figure 3 shows an example of this observation in 1997 chamber prototypes.

[1] 1 year = $10^7$ s is assumed. This estimate was done for the gas mixture $CF_4/CH_4$ (80:20), where a total ionization of the order of 160 e/cm was assumed. The corresponding value is about a factor 1.6 lower for the finally chosen gas $Ar/CF_4/CH_4$ (65:30:5). However, some safety factor to account for the presence of heavily ionising particles is included.

Such a behavior of the currents can be explained by the Malter effect [19]. An insulating layer on the cathode surface inhibits the neutralization of positive ions arriving from the avalanche at the anode. This generates an ion layer which accumulates and generates a strong electric field across the insulating layer. At some point the field becomes strong enough to extract electrons out of the cathode with enough energy to pass the ion layer, generating an avalanche and thus creating a self-sustaining current.

Several ad-hoc remedies were tested, such as adding about 1 percent water or alcohol to the gas. The water addition did not have any effect. Alcohol had prevented the self-sustaining currents to appear. However, the foils suffered mechanical deformations, and therefore this method was not a viable solution.

## 5. Reproduction of Malter currents

The restricted access to the HERA-B experiment made it difficult to perform a systematic R&D study of this effect in HERA. Since the Malter currents could never be observed under X-ray irradiation, it was necessary to find a radiation source that resembled the HERA-B conditions and was able to reproduce the Malter currents in chambers built as those installed in HERA-B.

Several parameters, like irradiation area, irradiation density and particle type were varied. Table 2 shows a summary of these tests. All chambers tested were built using the materials specified in Table 1. Chambers were typically 16 cells wide and 10-100 cm long. Several conclusions can be extracted from these tests:

- Electromagnetic radiation is not able to produce Malter currents. However, if the chamber has previously been irradiated in a hadronic beam (even if did not show Malter currents because of the low irradiation received), Malter currents could be observed. Once triggered, chambers are permanently damaged.
- Only irradiation induced by hadronic particles above certain energy clearly reproduce the Malter effect after a few mC/cm of accumulated irradiation dose, as it was observed in test modules installed in HERA-B.
- Slight indications that the irradiation area could also play a role were observed in the test using 70 MeV/c protons in PSI.

Table 2. Characteristics of the aging tests performed with irradiation sources aiming to reproduce the Malter currents in chambers built in the same way as the prototypes tested in HERA-B in 1997.

| Radiation Type (Facility) | Accum. Charge (mC/cm) | Radiation Density (µA/cm) | Irradiation Area (cm x cm) | Gas Mixture[a] | Malter Currents Seen? |
|---|---|---|---|---|---|
| X-rays Mo 35 kV (DESY Zeuthen) | 5000 | 1.5 | 1 x 3 | $CF_4$/$CH_4$ | NO[b] |
| X rays Cu 8 keV (Dubna) | 6000 | 5 | 0.5 x 1 | Ar/$CF_4$/$CO_2$ | NO[b] |
| Electrons 2.5 MeV (Hahn-Meitner Institut) | 10 | 0.1-3 | 100 x 30 | Ar/$CF_4$/$CH_4$ | NO[b] |
| X-rays Cu 8 keV (Univ. Heidelberg) | ~5-10 | 0.1 | 46 x 30 | Ar/$CF_4$/$CH_4$ | NO[b,c] |
| Protons 13 MeV/c (FZ-Rossendorf) | 5 | 0.3 | 9 x 9 | Ar/$CF_4$/$CH_4$ | NO |
| α-particles 28 MeV/c (FZ-Rossendorf) | 3 | 0.6 | 1 x 3 | Ar/$CF_4$/$CH_4$ | NO |
| Protons 70 MeV/c (Paul Scherrer Inst.) | ~5-10 | 0.2 | 0.5 x 0.5 | Ar/$CF_4$/$CH_4$ | NO[d] |
| π/protons 350 MeV/c (Paul Scherrer Inst.) | ~5-10 | 0.02 | 12 x 22 | $CF_4$/$CH_4$ | YES |
| α-particles 100 MeV/c (FZ-Karlsruhe) | ~5-10 | 0.4 | 7 x 7 | Ar/$CF_4$/$CH_4$ | YES[e] |
| pN Collisions (HERA-B) | ~5-10 | 0.02-0.04 | 100 x 30 | All gas mixt. | YES |

[a] Exact compositions as follows: $CF_4$/$CH_4$ (80:20), Ar/$CF_4$/$CH_4$ (74:20:6), Ar/$CF_4$/$CO_2$ (65:30:5)

[b] Malter effect currents could be triggered with this source in chambers that had been previously irradiated in hadronic beams able to trigger the Malter effect.

[c] A very strong anode aging and persistent dark currents were observed.
[d] Malter effect could be ignited by increasing the irradiation area to approximately 5 x 5 $cm^2$.
[e] Also seen with $CF_4/CH_4$ gas mixture.

The results in Table 2 suggest that a hadronic beam above a certain energy (about 100 MeV/c) is able to reproduce the aging effects seen in the HERA-B environment.

During these tests, it was also found that, in the 350 MeV/c $\pi$ beam at the Paul Scherrer Institute, a chamber where the Pokalon-C was coated with carbon spray (Graphit 33 Leitlack from Kontakt Chemie, CRC Industries Deutschland, Iffezheim) did not show Malter currents, thus supporting the hypothesis that the effect previously seen in HERA-B was caused by Malter currents produced by an insulating layer on the cathode foil surface.

In addition, at the large-area irradiation X-ray source in Heidelberg, a chamber operated with $Ar/CF_4/CH_4$ (74:20:6), showed a very fast anode aging phenomenon. The monitoring anode currents dropped by 50% after 8 hours of irradiation, corresponding to about 5 mC/cm. This effect was very difficult to observe in HERA-B because of the high target rate fluctuations. Another chamber with a very large Kapton window (of 100x50 $cm^2$) showed persistent dark currents after an accumulated charge of the order of 8 mC/cm.

In order to study systematically all these effects and to decide on the materials and production techniques of the detector, it was decided to do systematic tests at the 100 MeV $\alpha$-beam at the FZ-Karlsruhe.

## 6. Systematic aging investigations and validation of chamber construction parameters in a 100 MeV $\alpha$-beam.

The validation of parameters for the construction of the HERA-B Outer Tracker and gas system was made at DESY-Zeuthen (X-rays, Mo 35 kV), Hamburg (X-rays, Cu 8 keV) and in the 100 MeV/c $\alpha$-beam in the Forschungszentrum Karlsruhe. Details of the two first setups are given in [20,21,22]. Here, the investigations made with the 100 MeV/c $\alpha$-beam are described, where most of the basic chamber parameters were fixed.

For these investigations, 16-cell chambers with one layer of 30 cm length and 5mm cell diameter were built. The cathode material used was Pokalon-C coated with different metals: Cr, Au, Cu/Au with thickness varying between 50 to 100 nm. The materials and design for the chambers were carefully selected such that only one parameter at a time (e.g.: the Cu/Be wire, irradiation of FR4 strips, solder tin, different glues, etc…) was tested. The default materials are listed in Table 3. Chambers with materials as specified in Table 1 were also tested as a cross check. All chambers were installed in identical gas boxes made of clean aluminium frames and Kapton windows. All gas seals were done using the glues specified in Table 3. Several chambers, where gas was introduced directly into the cells by using a modified end-piece and steel capillary construction, were also built. In those cases, the sealing of the end-piece and capillary construction was done with Torrseal 8030 from Varian.

The open-loop gas system used only electro-polished stainless steel tubing. The default gas flow was one gas box volume exchange per hour. By default, the gas inlet was on the bottom of the gas box and the gas outlet on the top. The chambers were not separately flushed. In order to test the possible influence of the gas flow the chambers having gas input capillaries were also operated with 10 and 0.1 volume exchange per hour. For the test of outgassing of contact materials used during mass-production (eg: the anticontact paste used in the templates), gas was flushed through a box containing large samples of the material under study before entering the irradiated detector.

Table 3. Materials used in the modules tested in Karlsruhe.

| Component | Material |
|---|---|
| Cathode | Pokalon-C coated with Cu(50 nm)/Au(40 nm)[a] in APVV GmbH, Essen, Germany |
| Anode wire | 25 µm Au/W from California Fine Wires |
| Signal transport wire | 75 µm Cu/Be from Little Falls Alloy |
| Supporting strips | FR4 with Sn coated solder pads |
| Supporting glue | STYCAST 1266 / Catalyst 9 from Emmerson & Cummings |
| Conductive glue | Silber Leitkleber 2025 from Epoxy Produkte GmbH or Traduct 2922 from Tracon |
| Solder tin | FLUITIN 1603 Sn60Pb DIN 1707 F-SW32 DIN 8516, 3.5 % flux from Küppers Metallwerk GmbH |
| End-pieces | NORYL |
| Gas box | Aluminium frames and Kapton windows[b] |
| Gas system | Electro polished Stainless Steel tubes |

[a] Coatings of Cr, Au and Cr between 50 nm and 100 nm were also tested.
[b] In some chambers, aluminized Mylar foil was glued on top of the Kapton foil to disentangle the effect that Kapton is transparent to water.

Anode currents were continuously monitored with a precision of 1 µA. An $^{55}Fe$ spectrum of the irradiated wires was recorded periodically. The gas system monitored main (Ar, $CF_4$, $CH_4$ and $CO_2$) and trace gas components ($O_2$, $H_2O$) mainly by use of a chromatograph.

The tests served to solve the aging phenomena observed and to validate the final selection of chamber materials and production techniques.

### *6.1. Fast Anode Aging*

The fast anode aging observed when chambers were operated with the $Ar/CF_4/CH_4$ (74:20:6) gas mixture was confirmed in these tests. This aging effect was independent of the materials used and of the type of cathode foil. The $^{55}Fe$ spectra in the irradiation areas showed a decrease in gain compatible with the anode current drop observed. The curve in fig. 4 shows the $^{55}Fe$ spectrum along one wire after an accumulated charge of 5 mC/cm. The region where the $^{55}Fe$ peak position is lower corresponds to the irradiation area. The inspection of damaged chambers showed deposits in the form of fragile whiskers on the anode wires. By means of Energy Dispersive X-ray Spectroscopy (EDS) in combination with Scanning Electron Microscopy (SEM), it was shown that the main constituents were C and F in varying proportions, sometimes accompanied by Si.

None of the chambers operated with $Ar/CF_4/CO_2$ (65:30:5) gas mixture showed any sign of anode aging. Tests done with $Ar/CO_2$ (80:20), a gas too slow for use in HERA-B, did not show any anode aging either.

Moreover, a chamber previously aged using the $CH_4$-based mixture, was irradiated for some hours after changing the gas to $Ar/CF_4/CO_2$ (65:30:5). The points on fig. 4 show the peak position of the $^{55}Fe$ spectrum along one wire after irradiation with the $CO_2$-based gas mixture. One can clearly see how the chamber is completely recovered after exchanging the gas. Consequently, for the operation of the OTR in HERA-B, the $Ar/CF_4/CO_2$ (65:30:5) gas mixture was chosen.

*6.2. Malter effect*

Malter currents were observed for all chambers made with uncoated Pokalon-C and operated with $Ar/CF_4/CH_4$ (74:20:6). All these chambers previously showed massive anode aging as explained in section 6.1. Inspection of the cathode foil by electron microscopy showed the following [23]:

- Using SEM, it was found that for low electron energies, the surface of non-aged Pokalon-C foil behaved like an insulating layer. At a transition energy of about 0.9 keV, which probes deeper into the material, the foil surface looked conductive. This observation could be interpreted as the foil having an insulating layer of about 100 nm, which may be a result of the production procedure.
- The inspection of cathode foil from aged chambers with electron-emission spectroscopy for chemical analysis showed CF, CO and CN components in a polymerized form attached to the foil. The nitrogen content in the spectra could be an indication for outgassing of the epoxy used for gluing some of the test chambers that were built like the first prototype (Araldite AW 106, hardener contains CN components).

Chambers built with uncoated Pokalon-C operated with $Ar/CF_4/CO_2$ (65:30:5) did not show clear Malter currents. However, in some cases, spurious rest currents and instabilities were observed both in the 100 MeV/c $\alpha$-beam and in HERA-B. None of the chambers built with Pokalon-C with coating of any kind showed Malter effect, independently of the gas mixture used and of whether they showed anode aging or not. Clearly, any metal coating provides sufficient surface conductivity for the cathode to avoid completely the Malter effect. For the construction of the OTR it was consequently decided to use Pokalon-C coated with 40 nm Au on top of 50 nm Cu. This specific choice was driven by the fact that Cu attaches better to plastics than Au. Additionally, non-outgassing glues were exclusively used and the material cleaning procedures were made stricter for mass production [24].

*6.3. Persistent dark currents*

After an irradiation dose of 300 mC/cm, the chamber produced with Pokalon-C coated with Cu(50 nm)/Au(40 nm) and operated with $Ar/CF_4/CO_2$ (65:30:5) gas mixture at 1/10 of the nominal volume exchange in the drift tubes (using the special capillari construction) started to draw currents with ohmic behaviour, similar to those observed in the Heidelberg X-ray tests. It was found that these currents were caused by the supporting FR4 strips, which became conductive during irradiation. Inspection of the strips showed strong corrosion both of the epoxy of the strips (the glass fiber texture became quite pronounced) and of the solder points. It was further found out that the water content of the gas had exceeded 3000 ppm during the irradiation period, due to the Kapton windows used in the gas boxes and the low operation gas flow.

A systematic test varying the water concentration of the gas was performed. To avoid uncontrolled sources of water due to the transparency of the Kapton windows, an additional aluminized Mylar foil was glued to the gas windows. The chamber was operated at the nominal gas flow and water was added to the gas by flushing it through water vapor in a temperature-controlled water bath. Figure 6 shows the measurement of dark currents versus the water concentration. An almost linear dependency of the ohmic dark current with the water concentration is observed. However, ohmic currents do not appear if water concentrations are below about 500 ppm.

This fact clearly has implications for the gas system and the gas box construction of the final HERA-B Outer Tracker detector that have to ensure enough gas-tightness to meet this requirement [18].

*6.4. Anode swelling and anode etching*

Figure 6 shows the peak positions of the $^{55}Fe$ spectrum along the wire for several wires irradiated in the 100 MeV $\alpha$-beam at FZ Karlsruhe after an accumulated charge of 600 mC/cm for chambers operated with $Ar/CF_4/CO_2$ (65:30:5). A slight 6-10 % gain drop is observed for wires in the center of the irradiation zone. Inspection of the anode wires showed that they had become slightly thicker (consistent with the gain drop) and that the gold partly had peeled off in some areas. Further laboratory tests [20] with irradiation levels between 0.4 and 0.7 μA/cm reproduced the effect if the water concentration was below about 100 ppm. However, another test [22] in which the irradiation level was 0.2 μA/cm and the water concentration was below 50 ppm did not confirm this result. Difficulties in measuring such low water concentrations as the reason for this discrepancy cannot be exluded.

**7. Summary**

The Outer Tracker of HERA-B has been constructed with honeycomb drift tubes. The harsh operating environment in terms of radiation dose and particle flux, which is very close to those expected at the LHC, required production methods of a new generation of gaseous detectors.

Aging studies are usually performed with small prototypes of the final detector. Special care has to be taken to the scalability, the influence of outgassing materials (which might be more abundant in the final system), the radiation density (that has to be balanced with the duration of the test) and the particle type. In particular, some of the aging effects produced in the final detector conditions for the HERA-B Outer Tracker could only be reproduced with a hadronic beam above a certain energy and, possibly, above certain irradiation area. The choice of the irradiation source for aging tests must also be a part of the detector R&D.

In the detector construction, several lessons have been learned from the extensive R&D done for the HERA-B Outer Tracker:

- Some aging effects might be particle type or energy dependent. Beam tests for aging studies must resemble running conditions of the final detector.
- Special care has to be taken of the surface conductivity of the plastic cathode foil. Microscopic non-conductive layers can be crucial for the safe operation of the detector. The aging effects caused by these areas might not appear in common X-ray laboratory tests and, as in the case of the HERA-B Outer Tracker, can appear after a very low irradiation dose.
- Methane can cause severe anode aging also in $CF_4$-based gases. The use of $CO_2$ as quencher was proven to be safe for the Outer Tracker, although at the price of reduced drift velocity.
- Special care has to be taken of the selection and cleaning procedures of all the materials that might be in contact with the gas. Non-outgassing components and careful control of material handling during production are mandatory.
- $CF_4$ is an expensive gas that forces large detectors to use a re-circulating gas system. The operation parameters of these systems have to be thoroughly tested in the detector aging studies. In particular, the water content and the gas flow can play a decisive role in the operability of the final detector.

The main lesson to successfully build and operate the new generation of gaseous detectors that can stand radiation doses on the order of 500 mC/cm per year is to make very systematic aging tests with prototypes containing all materials and building techniques in an environment resembling the final running conditions as closely as possible before starting production.

**Acknowledgements**

We thank S.Rolle for continuous support of the SEM investigations at the Technische Fachhochschule Wildau. We also thank P. Fehsenfeld and R. Kubat from HZY, Forschungszentrum Karlsruhe for their patient and friendly help in providing us with the 100 MeV/c α-beam. M.Cook, A.Hamel and A.Harrison were extremely valuable in collecting the data in the aging tests at Karlsruhe. We would finally want to thank A.Fee, J.Hansen, M.Hoffmann, M.Jablonski, J.Schulz, M.Wagner for the technical support without which the measurements would have not been possible.

**References**

[1] H. Hartouni et al., HERA-B Design Report, DESY-PRC-95/01, (1995).
[2] E. Gerndt, S. Xella, Nucl. Instr. and Meth. A 446 (2000) 264.
[3] H. Fleckenstein et al., Proc. IEEE Nucl. Sci. Symp., San Diego (2001).
[4] W. Wagner, Nucl. Instr. and Meth. A 446 (2000) 222.
[5] T. Zeuner, Nucl. Instr. and Meth. A 446 (2000) 324.
[6] A. Zoccoli, Nucl. Instr. and Meth. A 446 (2000) 246.
[7] M. Titov, Nucl. Instr. and Meth. A 446 (2000) 366.
[8] J. Pyrlik, Nucl. Instr. and Meth. A 446 (2000) 299.
[9] T. Hott, these proceedings.
[10] M. Hildebrandt, these proceedings.
[11] M. Titov, these proceedings.
[12] C. Stegmann, Nucl. Instr. and Meth. A 453 (2000) 153.
[13] H. Kolanoski, Proc. IEEE Nucl. Sci. Symp., Lyon (2000), Conference Record, 5-6 (2001).
[14] M. Hohlmann, Nucl. Instr. and Meth. A 461 (2001) 21.
[15] M. Capeans, C. Padilla, C. Stegman, “Module Production Guide”, HERA-B internal note (December 1998).
[16] K. Berkhan et al., Proc. Of 5$^{th}$ Snowmass Workshop on Electronics for LHC (September 1999).
[17] R. Zimmermann, PhD Thesis, Rostock University (Rostock, Germany)(1999).
[18] M. Hohlmann, these proceedings.
[19] L. Malter, Phys. Rev. 50 (1936).
[20] A. Schreiner, these proceedings.
[21] A..Schreiner, PhD Thesis, Humboldt University (Berlin, Germany) (2001).
[22] K. Dehmelt, these procedings.
[23] G. Bohm, these proceedings
[24] M. Capeans, these proceedings.

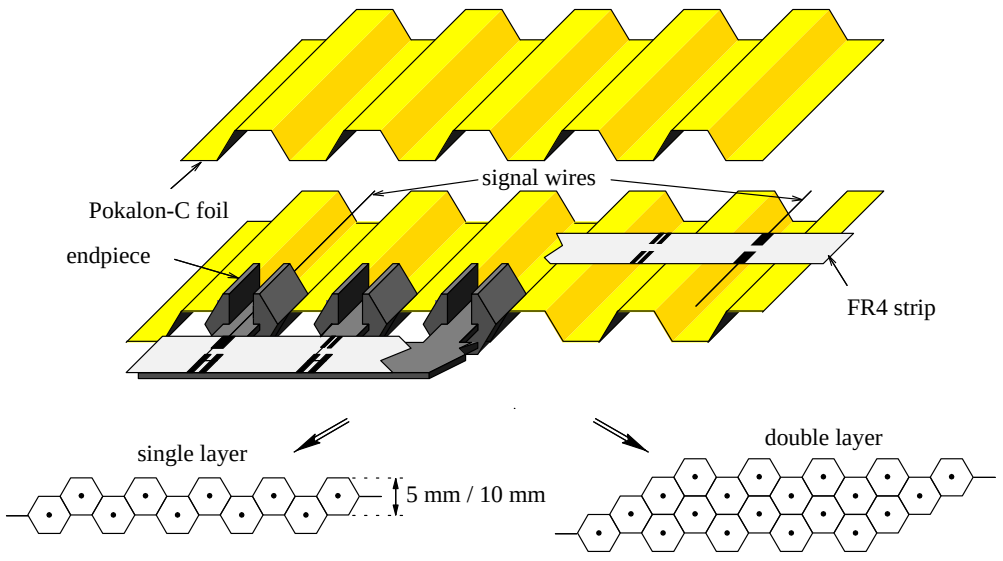

Fig. 1. Illustration of the honeycomb technology. See explanations of the building technique in the text. In the lower part of the figures, the cross section of single and double layers modules are shown.

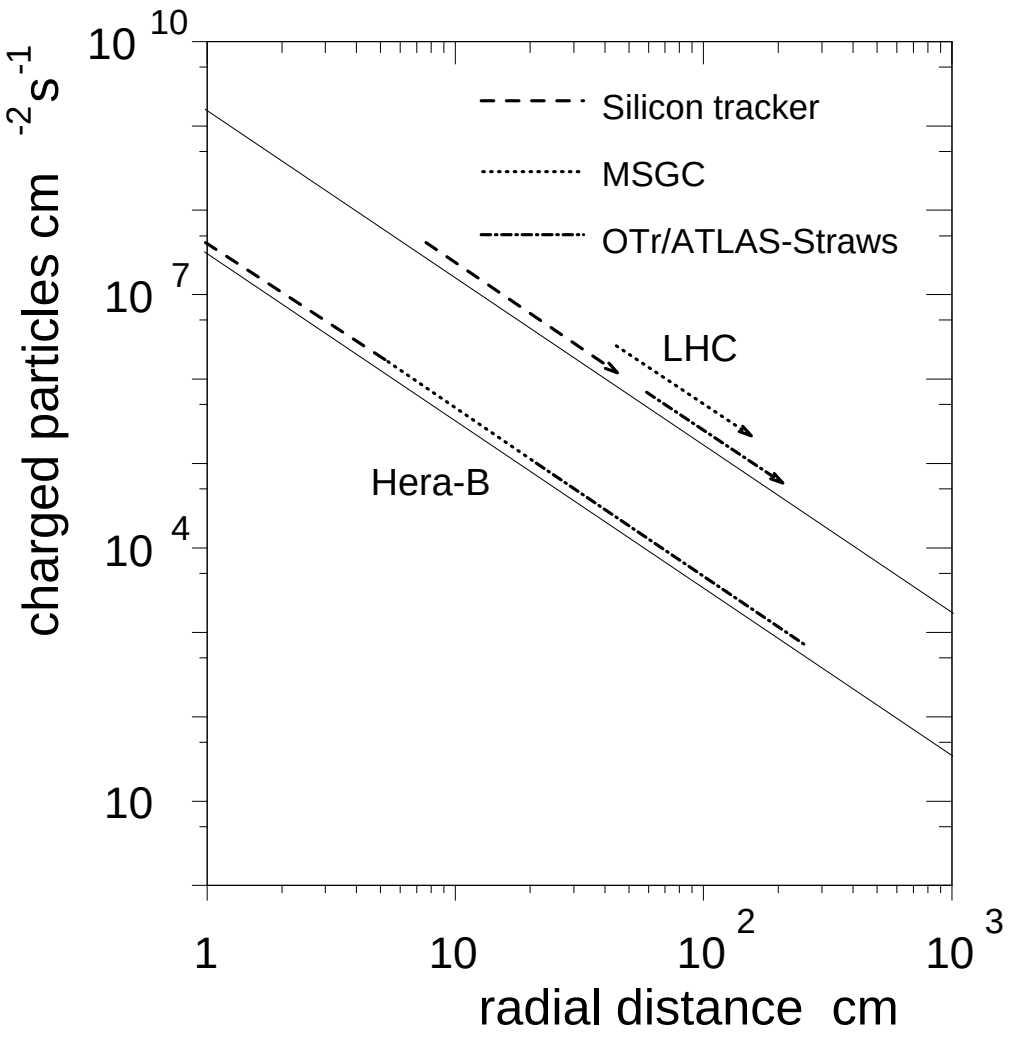

Fig. 2. Comparison of the primary charged particle density as function of the radial distance from the beam for detectors at HERA-B and LHC. It can be seen that, at HERA-B, detectors are closer to the beam, thus receiving the same charged particle flux, as LHC detectors will detect.

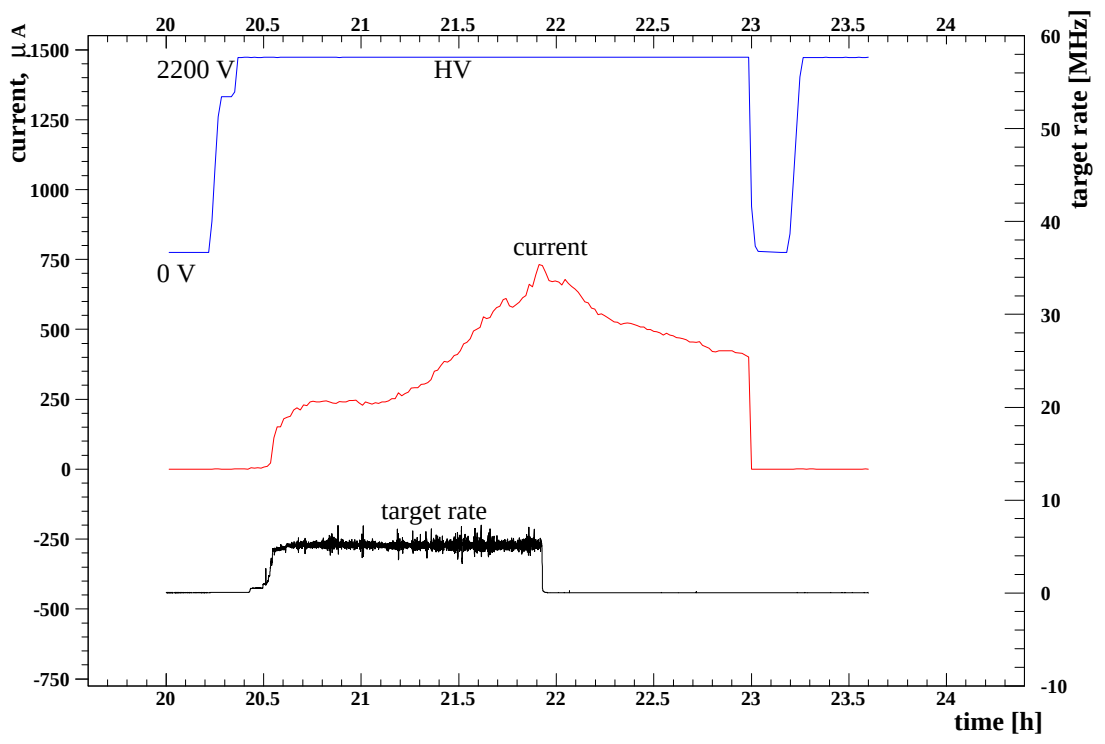

Fig 3. Example of the development of Malter currents in 1997 test chambers installed in HERA-B. The chambers were built with uncoated Pokalon-C foil. The drift gas was $CF_4/CH_4$ (80:20). The figure shows the time behavior of the high-voltage applied to the anode wire, the anode current measured and the target rate (which is proportional to the radiation load in the chambers).

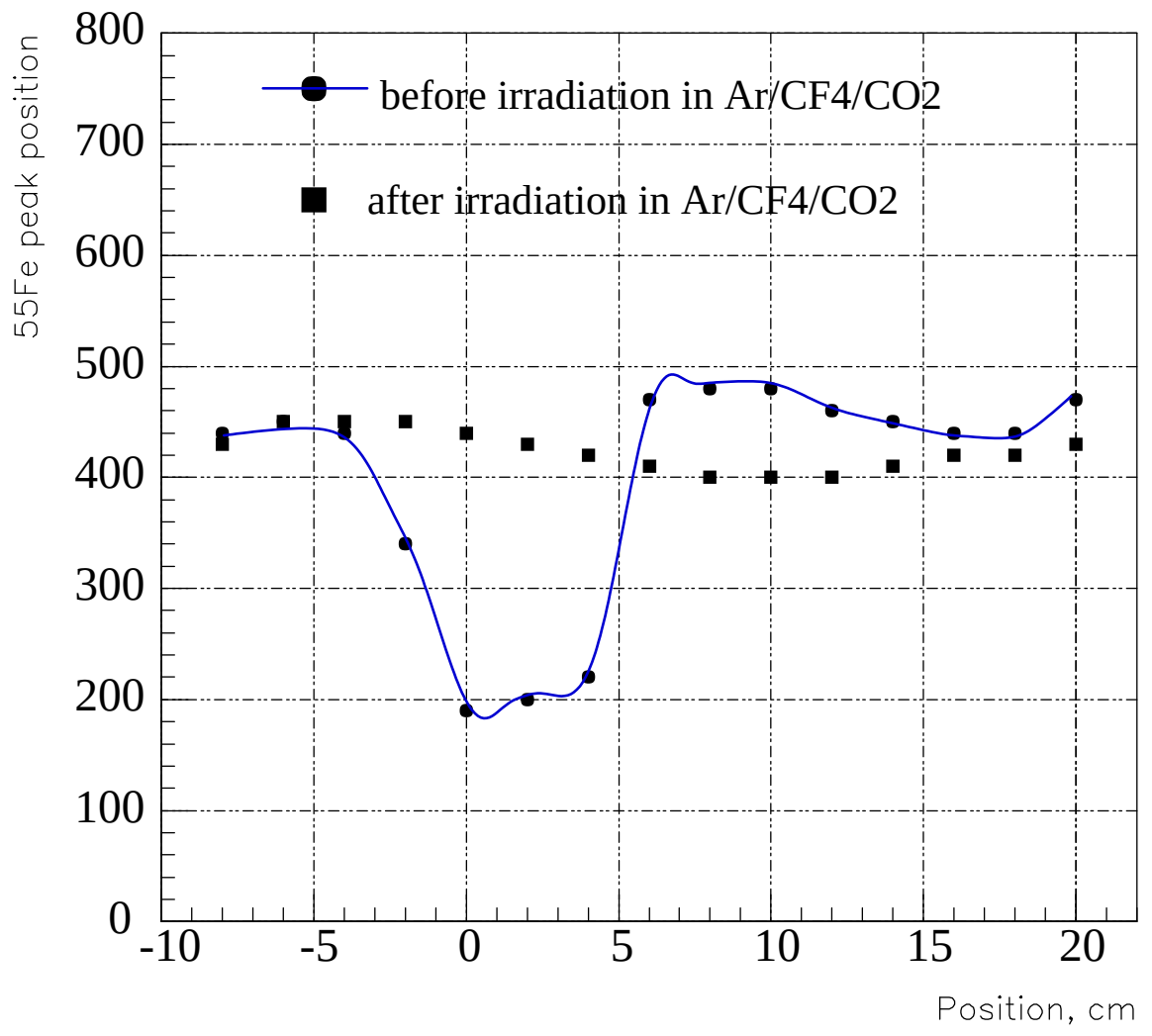

Fig. 4. Position of the $^{55}Fe$ peak as function of the source position along the wire after severe anode aging produced by irradiation under $Ar/CF_4/CH_4$ (74:20:6) (full circles and curve) and after recovering from irradiation by operating with $Ar/CF_4/CO_2$ (65:30:5) (squares).

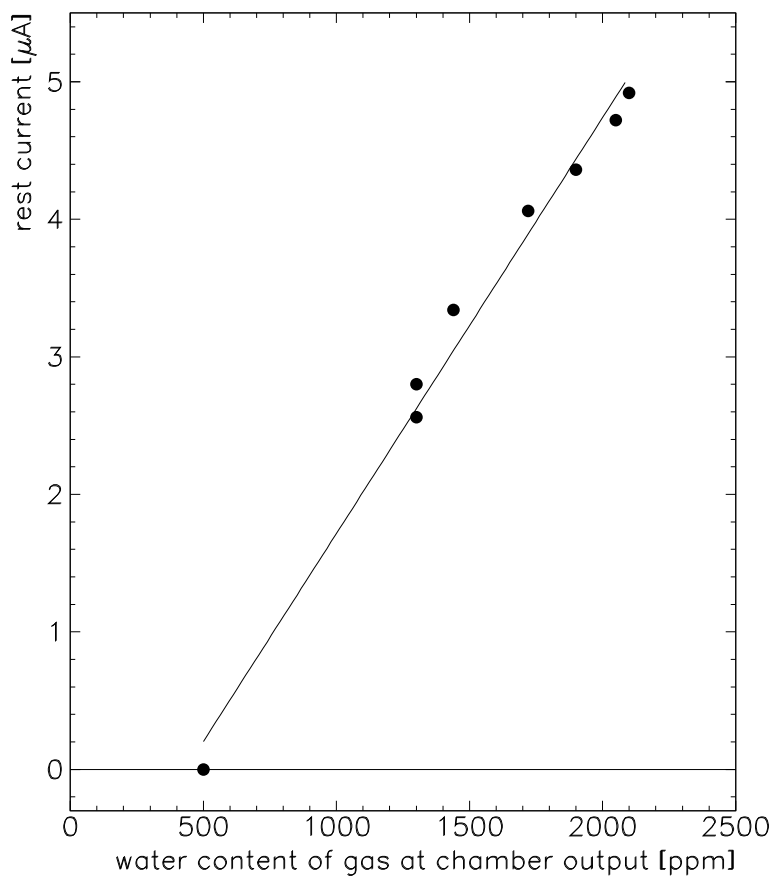

Fig. 5. Measurement of dark currents versus the water concentration. An almost linear dependency of the ohmic dark current with the water concentration is observed. The gas used is $Ar/CF_4/CO_2$ (65:30:5).

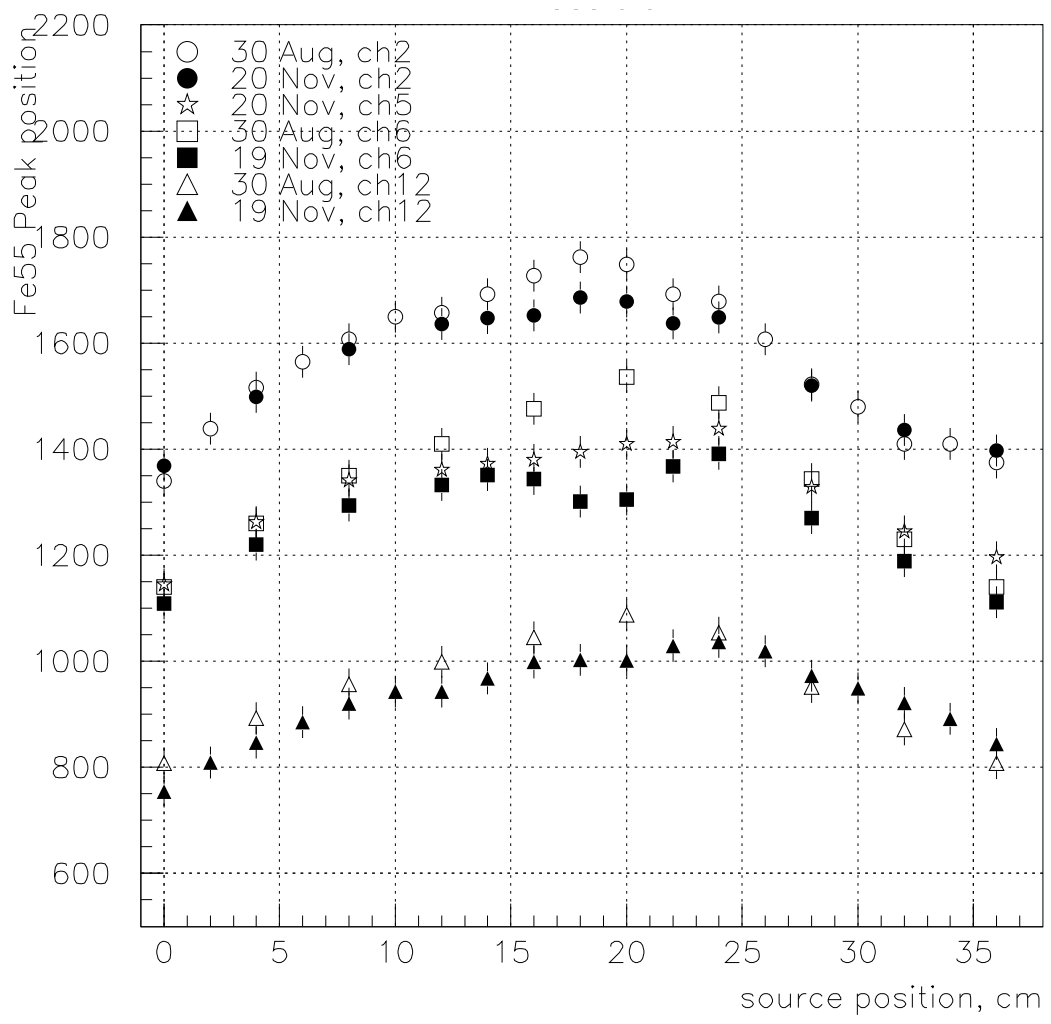

Fig. 6. Peak position of the $^{55}$Fe spectrum along the wire for several irradiated wires after an accumulated charge of 600 mC/cm in chambers operated with $Ar/CF_4/CO_2$ (65:30:5) (full symbols), compared with the situation before irradiation (open symbols). The gain drop is between 6-10 % in the center of the irradiation zone due to the swelling of the anode wires.